\documentclass[fleqn,10pt]{wlscirep}
\usepackage[utf8]{inputenc}
\usepackage[T1]{fontenc}

\usepackage{comment}
\usepackage{microtype}
\usepackage{graphicx} 
\usepackage{titlesec} 
\usepackage{url}
\usepackage{multicol,lipsum} 

\usepackage{csquotes} 
\usepackage{setspace} 
\usepackage{hyperref} 
\usepackage{pdfpages} 
\usepackage{multirow}
\usepackage{datetime}
\usepackage{adjustbox}
\usepackage{float}
\usepackage{longtable}
\usepackage{array}
\usepackage{xcolor}

\title{MDF-Net for Abnormality Detection by Fusing X-Rays with Clinical Data}

\author[1]{Chihcheng Hsieh}
\author[2,+]{Isabel Blanco Nobre}
\author[2,+]{Sandra Costa Sousa}
\author[1]{Chun Ouyang}
\author[1]{Margot Brereton}
\author[3]{Jacinto C. Nascimento}
\author[4]{Joaquim Jorge}
\author[4,5,*]{Catarina Moreira}
\affil[1]{Queensland University of Technology, Brisbane, Australia}
\affil[2]{Lus\'{i}adas Knowledge Center, Imageology Department, Lisbon, Portugal}
\affil[3]{Instituto de Sistemas e Rob\'{o}tica, Lisbon, Portugal}
\affil[4]{Instituto Superior T\'{e}cnico, University of Lisbon, Portugal / INESC-ID}
\affil[5]{Data61 CSIRO, Eveleigh, NSW, Australia}
\affil[*]{catarina.pintomoreira@data61.csiro.au}

\affil[+]{these authors contributed equally to this work}

\keywords{Deep Learning, Multimodal Learning, Fusion Methods, Object localisation, Chest X-rays}

\begin{abstract}
This study investigates the effects of including patients' clinical information on the performance of deep learning (DL) classifiers for disease location in chest X-ray images. Although current classifiers achieve high performance using chest X-ray images alone, our interviews with radiologists indicate that clinical data is highly informative and essential for interpreting images and making proper diagnoses.

In this work, we propose a novel architecture consisting of two fusion methods that enable the model to simultaneously process patients' clinical data (structured data) and chest X-rays (image data). Since these data modalities are in different dimensional spaces, we propose a spatial arrangement strategy, 
{\it spatialization}, to facilitate the multimodal learning process in a Mask R-CNN model. We performed an extensive experimental evaluation using MIMIC-Eye, a dataset comprising modalities: MIMIC-CXR (chest X-ray images), MIMIC IV-ED (patients' clinical data), and REFLACX (annotations of disease locations in chest X-rays).

Results show that incorporating patients' clinical data in a DL model together with the proposed fusion methods improves the disease localization in chest X-rays by 12\% in terms of Average Precision compared to a standard Mask R-CNN using only chest X-rays. Further ablation studies also emphasize the importance of multimodal DL architectures and the incorporation of patients' clinical data in disease localization.  
The architecture proposed in this work is publicly available to promote the scientific reproducibility of our study (\url{https://github.com/ChihchengHsieh/multimodal-abnormalities-detection}).
\end{abstract}
\begin{document}

\maketitle

\section*{Introduction}
According to the Lancet~\cite{Haakenstad2022}, 2019 witnessed a shortage of $6.4$ million physicians, $30.6$ million nurses, and $2.9$ million  pharmaceutics personnel across $132$ countries worldwide, especially in Low and Medium Income Countries. The situation has worsened after the pandemic since medical staff were disproportionately affected.

Deep Learning (DL) technologies promise to deliver benefits for health systems, professionals, and the public, making existing clinical and administrative processes more effective, efficient, and equitable. These technologies have become highly popular in the medical imaging field, where a plethora of applications have been successfully addressed e.g., breast imaging \cite{Maicas19,Li2019}, left ventricular assessment \cite{Liu2020,Medley2022}, dermoscopy analysis \cite{Pham2021,Haenssle2018} and chest X-Rays \cite{Irvin2019Chexpert,Rajpurkar2017CheXNet,Rajpurkar2018CheXNeXt,Yates2018RedDotCXR}, which have gained further attention  due to the recent pandemic.
Despite their advantages, these systems have notable shortcomings; they require extensive amounts of labelled data to operate correctly. Explicitly labelling anomalies in large amounts of medical images requires the availability of medical experts, who are scarce and expensive. The process is time-consuming and costly, resulting in bottlenecks in research advancements \cite{Rahimi21}. Additionally, the complex interconnected architectures make DL predictions opaque and resistant to scrutiny, hindering the adoption of AI systems in public health (known as the ``black box'' problem~\cite{Guidotti2018,Lipton2018,Miller2019,Moreira2021DSS,Moreira2020icsoc,Moreira2022fusion,velmurugan2021evaluating,wickramanayake2022building}). A system that could automatically annotate or highlight relevant regions in medical images in a similar way to humans would be extremely useful and could save millions of dollars creating breakthroughs in research in AI adoption in Healthcare \cite{Rahimi21}.

Several works in the literature attempt to automatically learn regions of interest indicating the patient's clinical abnormalities. These works mainly use DL approaches which have been found to be efficient and effective on a variety of computer vision tasks \cite{Singh2020ExplainableMedicalImage,Zhuang2019TransferLearningSurvey,Yuan2020DAM,Tang2020AbnormalityCXRCNN}. Mask R-CNN \cite{Kaiming2017MaskRCNN} is one of the most widely used DL architectures to predict regions with abnormalities in images. 
However, their predictive performance is still low and these architectures do not take into consideration the process of how expert radiologists assess and diagnose these images. The human component is completely disregarded in most DL studies to predict abnormalities in chest X-rays. This is relevant because when radiologists look at an X-ray image they experience it in a multimodal world: they see objects, textures, shapes, etc. It is the combination of these modalities that make humans capable of making mental models that generalize well with less data when compared to DL approaches.

Advancements in computer-aided diagnostics (CAD) employing radiomics and deep learning have also been explored in the literature. A compelling example can be seen in a study that presents an attention-augmented Wasserstein generative adversarial network (AA-WGAN) for fundus retinal vessel segmentation. The application of attention-augmented convolution and squeeze-excitation modules highlights regions of interest and suppresses extraneous information in the images, proving effective in segmenting intricate vascular structures \cite{Liu23Attention}. Further, an attention-based glioma grading network (AGGN) for MRI data shows superior performance, highlighting the key modalities and locations in the feature maps even without manually labelled tumour masks \cite{Wu23AGGN}. Lastly, a CAD model, Cov-Net, exhibits robust feature learning for accurate COVID-19 detection from chest X-ray images, outperforming other computer vision algorithms \cite{Li22covnet}. Together, these studies underscore the potential of radiomics and deep learning in improved health anomaly detection, segmentation, and grading.

Multimodal DL consists of architectures that can learn, process, and link information from different data modalities (such as text, images, structured data, etc) \cite{Bayoudh2021MultimodalSurvey,Wang2020DeepMultimodal,Badgeley2019MultimodalHipFracture,Lahat2015MultiModalDataFusion,Ramachandram2017DeepMultiModalSurvey,Smilkov2017SmoothGrad}. Deep learning can benefit from multimodal data in terms of generalization and performance compared to the unimodal paradigm (see for instance Azam et al.\cite{Azam22review} for a comprehensive review in medical multimodal images). In terms of multimodal DL approaches for chest X-ray images, most works in the literature focus on combining image data with text to generate reports, predict diseases or even for lesion detection by training BERT-like models \cite{Moon22, Yan2022,Chen22, Rayner2019}. However, in terms of disease classification, the medical reports associated with the images that are used for training already contain some information about the patient's diseases, which may generate biased results.

A recent literature review~\cite{Castillo21} indicated that clinical data is highly informative and essential for radiologists to interpret and make proper diagnoses. To the best of our knowledge, there are no multimodal DL approaches that combine patients' clinical information with X-ray images to predict the location of abnormalities in chest X-rays. This justifies and motivates our research path in the present paper. Concretely, chest X-ray images and clinical data are aimed at addressing a critical gap in current object detection deep learning approaches in terms of fusing tabular data and image data, which are very scarce. Our interviews with radiologists reinforced the importance of clinical data in making accurate diagnoses from chest X-ray images because radiologists cannot make an accurate assessment of an X-ray image without knowing the patient’s clinical data. By integrating these two types of data into our model, we aim to capture the nuanced decision-making process of radiologists more effectively. This approach allows the model to leverage not just the patterns visible in the images, but also the rich contextual information available in the clinical data, leading to a more holistic and accurate disease localization.

\begin{figure}[!h]
    \centering
    \includegraphics[width=\columnwidth]{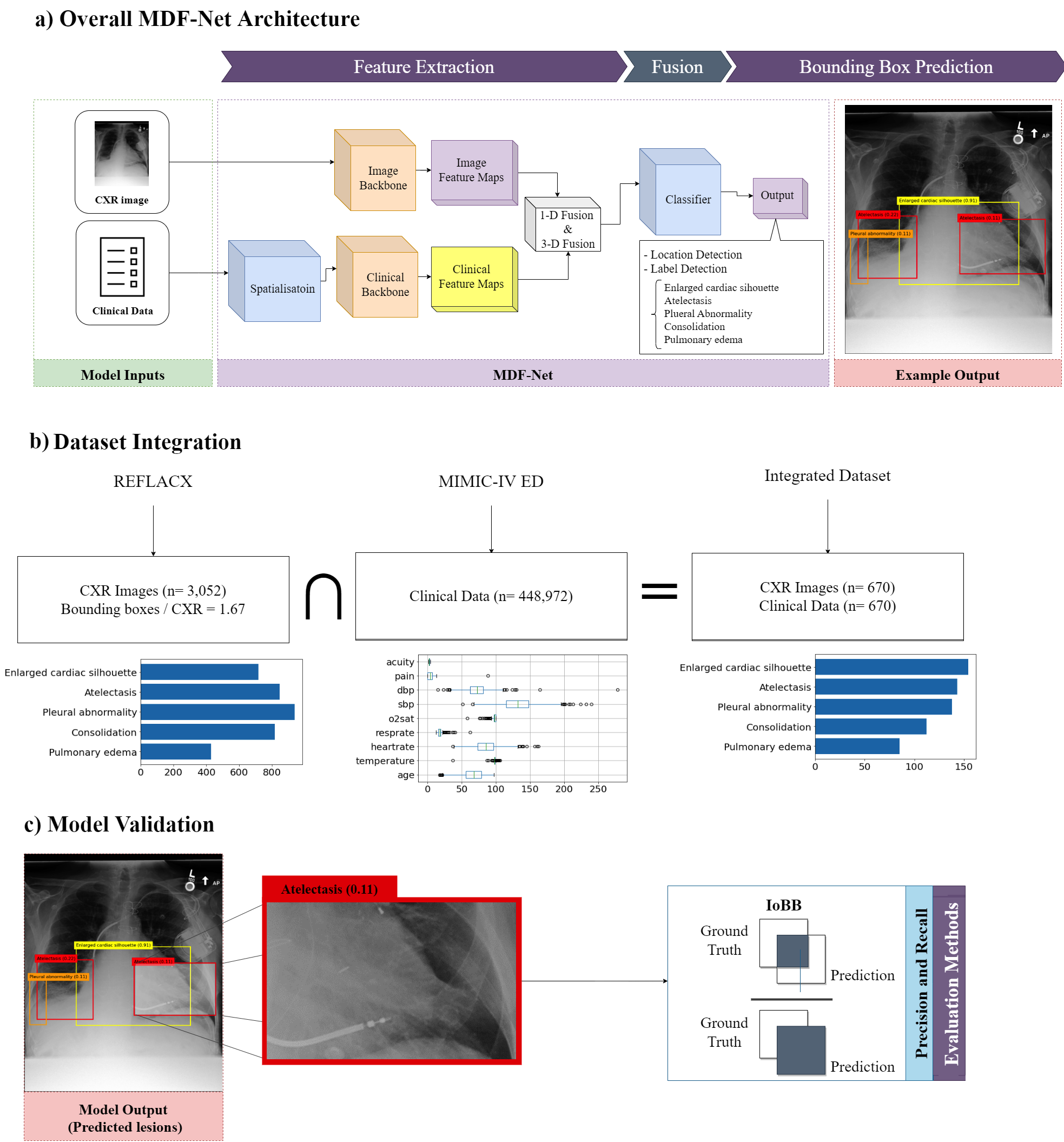}
    \centering
    \caption{Panel \textbf{a)} An overview of the proposed MDF-Net architecture. Panel \textbf{b)} Integration of the CXR images from MIMIC-CXR \cite{DJohnson2019MIMIC_CXR_JPG} with the clinical data of MIMIC IV \cite{Johnson2021MIMIC_IV}. Panel \textbf{c)} IoBB is used to evaluate the model between groundtruth and predictions.}
    \label{fig:method-overview}
\end{figure}

In this paper, we propose the \textit{Multimodal Dual-Fusion Network (MDF-Net)}, which is a novel architecture inspired and extended from Mask R-CNN~\cite{Kaiming2017MaskRCNN}. MDF-Net can fuse chest X-ray images and clinical features simultaneously to detect regions in chest X-rays with abnormalities more accurately.
The proposed architecture uses a two-stage detector comprising a Region Proposal Network (RPN) of Mask R-CNN followed  by an attention mechanism to extract information only from Regions of Interest (RoIs).  Figure~\ref{fig:method-overview} presents a general description of the proposed framework. Figure \ref{fig:method-overview} panel \textbf{a)} shows the overall architecture of the proposed model. The prediction process can be divided into three phases. 
The first phase aims to extract the semantics (feature maps) from input data. The two modalities are processed separately. One branch computes a feature map from the input images, and a second branch computes a  feature map from clinical data. In the second phase, we conduct a fusion operation to fuse the two feature maps above to obtain a joint representation. Then, the third phase applies the final classifier to predict the bounding boxes of abnormality. In Figure \ref{fig:method-overview} panel \textbf{b)}, we integrated the $triage$ data from MIMIC-IV ED with REFLACX in order to get corresponding clinical data for each CXR image. The integrated (multimodal) dataset allows us to perform multimodal learning with the model shown in Figure \ref{fig:method-overview} panel \textbf{a)}. In the end, we performed two ablation studies: one that investigated the impact of the different fusion methods in our architecture; and another that investigated the impact of different sets of clinical features for abnormality detection. The performance of the models was measured using Average Precision (AP) and Intersection Over predicted Bounding Boxes (IoBB) (Figure~\ref{fig:method-overview} panel~\textbf{c}). 

The key contributions of this work are as follows:
    (1) We propose a strategy to extract corresponding clinical data for CXR images from the MIMIC \cite{Johnson2019MIMIC_CXR,Johnson2021MIMIC_IV_ED,Johnson2021MIMIC_IV} dataset. This strategy is then used to construct our own multimodal dataset for the abnormality detection task;
    (2) We propose a multimodal learning architecture and two fusion methods to fuse tabular and image data together;
    (3) We demonstrate the effectiveness and importance of clinical data in abnormality detection  through ablation studies.

\begin{figure}[!ht]
    \centering
    \includegraphics[width=1\columnwidth]{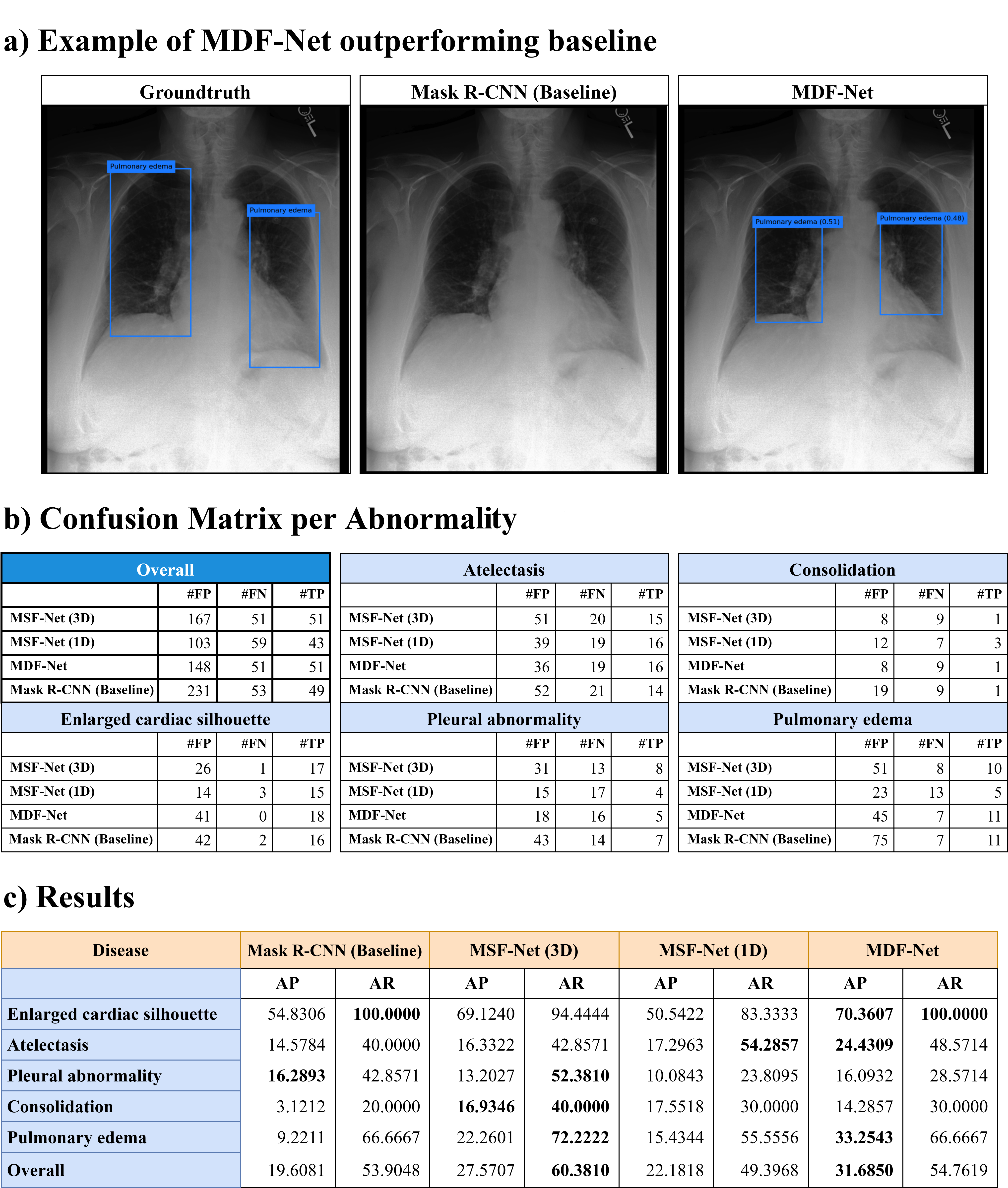}
    \centering
    \caption{Panel~\textbf{a)} The groundtruth and predictions from Mask R-CNN (Baseline) and MDF-Net. In this figure, MDF-Net picked up both pulmonary edema, which our radiologists mentioned the clinical can be helpful on identifying it. Panel \textbf{b)} Confusion matrix per abnormality (Note: The number of true negative cases is infinite in the object detection task). Panel~\textbf{c)} Evaluation results (Score threshold = 0.05, IoBB threshold = 0.5).}
    \label{fig:EvaluationFigure}
\end{figure}

\section*{Results}
 Overall, our experiments show that clinical data plays an important  role in abnormality detection. Using clinical data together with chest X-ray images, the proposed MDF-Net architecture achieved a significant improvement when compared to the baseline Mask R-CNN model using X-rays only. Figure \ref{fig:EvaluationFigure} panel \textbf{a)} shows an example where the proposed MDF-Net was able to correctly predict a bilateral pulmonary edema while the Mask R-CNN (baseline) did not identify any abnormality. In terms of the number of false negatives/false positives, the proposed MDF-Net with both fusion methods always generated fewer false positives and false negatives for the majority of the chest abnormalities that it was trained on (Figure~\ref{fig:EvaluationFigure} panel \textbf{b)}). 

 We also performed ablation experiments to evaluate the effectiveness of our fusion methods and the impact of different sets of clinical features. The implementation of this work is open-sourced for further research and reproducibility \cite{MDF-NET2022}. However, the dataset is restricted-access and requires users to fulfil PhysioNet's requirements to download and use it.

To evaluate the effectiveness of each fusion method proposed in our MDF-NET, we created two Multimodal Single Fusion Networks (MSF-Net), which only apply 1-D or 3-D fusion to conduct ablation studies. Figure \ref{fig:EvaluationFigure} panel \textbf{c)} presents the performance of Mask R-CNN (Baseline), MDF-Net and the two MSF-Nets. Figure \ref{fig:EvaluationIoBBs} panel \textbf{b)} presents the evaluation results across different IoBB thresholds.

\subsection*{Impact of Different Backbones}
\begin{figure}[!h]
    \centering
    \includegraphics[scale=0.6]{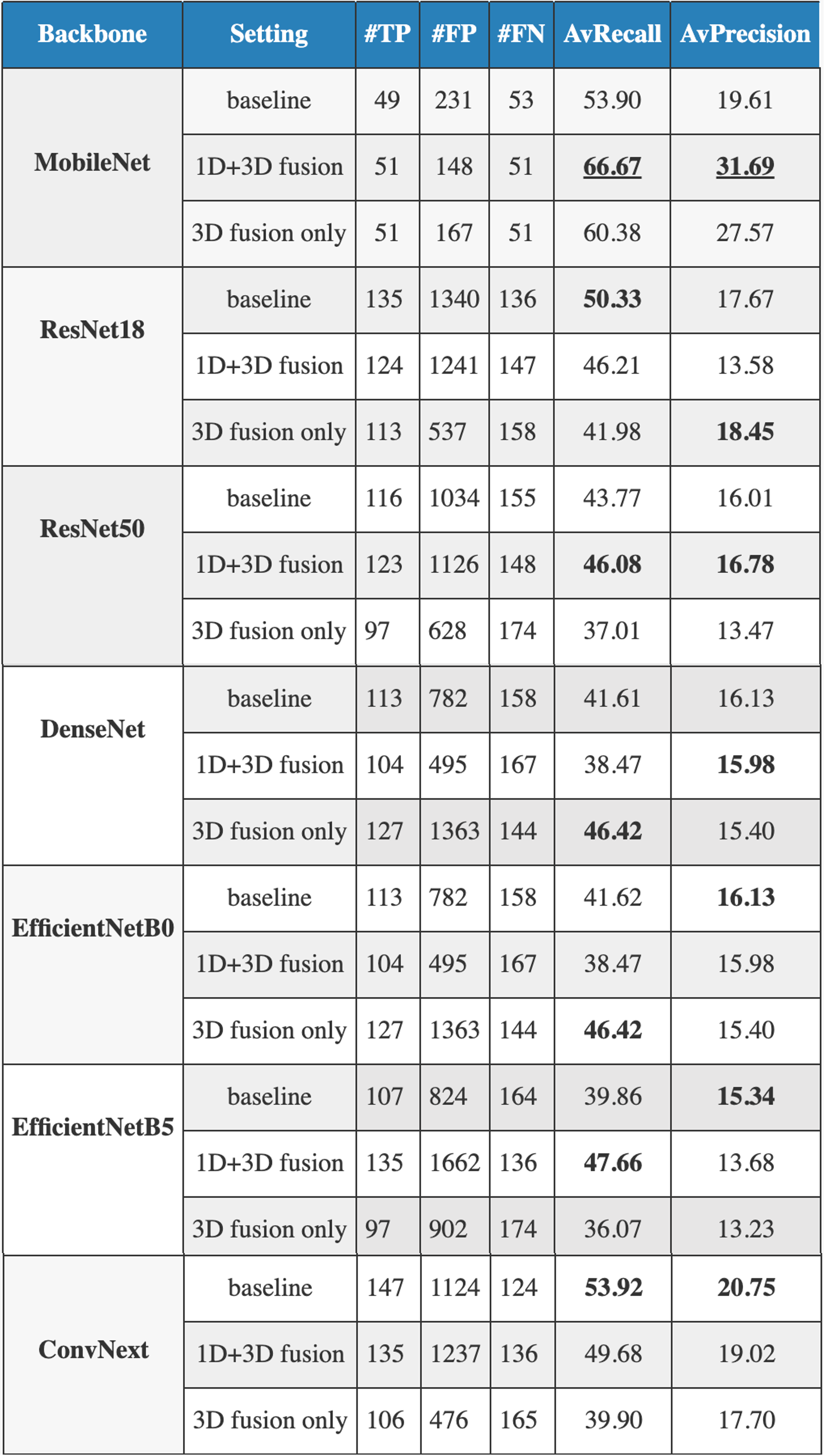}
    \caption{Ablation study results for different backbone architectures in the MDF-Net. This table provides the average precision (AP) and average recall (AR) values obtained using MobileNet, EfficientNet, ResNet, DenseNet, and ConvNextNet and also the overall number of True Positives (TP), False Positives (FP) and (False Negatives).}
    \label{fig:ablation_backbones}
\end{figure}

In order to thoroughly evaluate the performance of our proposed MDF-Net model and to study its robustness across different architectures, we conducted an ablation study focusing on different backbones. Figure \ref{fig:ablation_backbones} presents the obtained results. Backbone architectures play a pivotal role in deep learning models as they are responsible for the feature extraction process. For this study, we incorporated well-known architectures including MobileNet, EfficientNet, ResNet, DenseNet, and ConvNextNet, which exhibit diverse architectural designs. We systematically analyzed and compared their performance when incorporated as the backbone of our MDF-Net model for different settings: baseline (image only), using 3D fusion only and using both 3D and 1D fusion. This analysis aims to demonstrate the versatility of our proposed approach and to identify the backbone that can further enhance the performance of MDF-Net in disease localization tasks using chest X-rays and patients' clinical data. The ablation results indicate that MobileNet still outperforms all other backbones.

\subsection*{Impact of Different Fusion Methods}
\begin{figure}[!bh]
    \centering
    \includegraphics[scale=0.6]{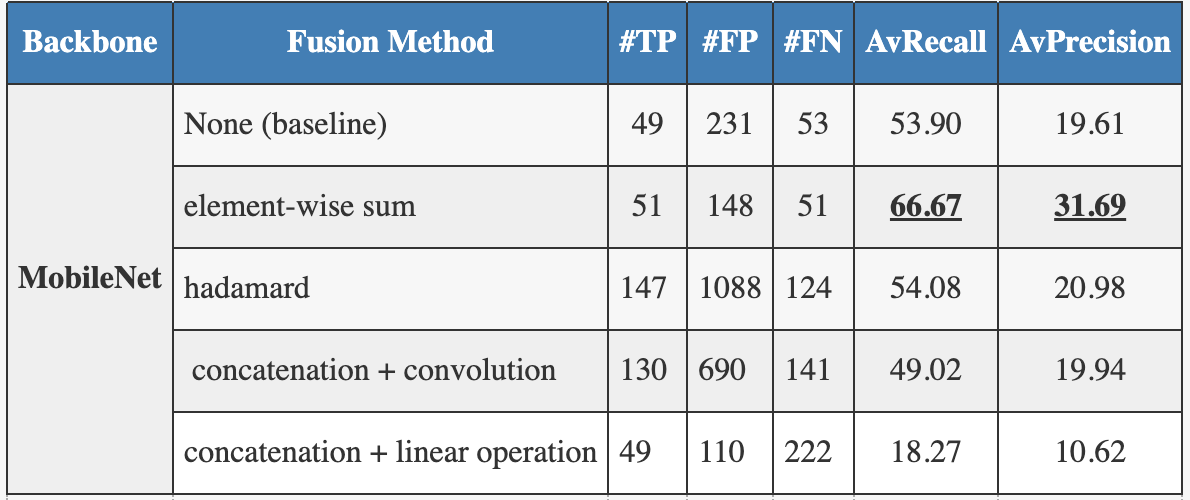}
    \caption{Ablation study results showcasing the impact of different fusion methods on the performance of the MDF-Net model. The table compares the performance metrics, namely, average precision and average recall, for the four fusion methods examined: element-wise sum, concatenation followed by a linear operation, concatenation followed by a convolution operation, and the Hadamard product.}
    \label{fig:ablation_fusion}
\end{figure}

Fusion methods play an instrumental role in multimodal deep learning models, acting as a bridge that intertwines the information derived from different data modalities. To scrutinize the effectiveness and compatibility of various fusion methods within our proposed MDF-Net, we conducted an ablation study where we tested different fusion strategies. The strategies assessed included element-wise sum, concatenation followed by a linear operation, concatenation followed by a convolution operation, and the Hadamard product. Each of these methods amalgamates information in distinct ways, carrying unique assumptions about the interplay between the features derived from the image and clinical data. The results of our ablation study illustrate that the element-wise sum fusion method yielded the best performance within our MDF-Net model. This method, which combines features by adding them together element by element, appeared to be more effective at integrating the information from chest X-ray images and clinical data, thus improving the model's ability to accurately localize disease in chest X-rays. Figure \ref{fig:ablation_fusion} presents our results.

\subsection*{Impact of Different Clinical Features}
\begin{figure}[!h]
    \centering
    \includegraphics[scale=0.4]{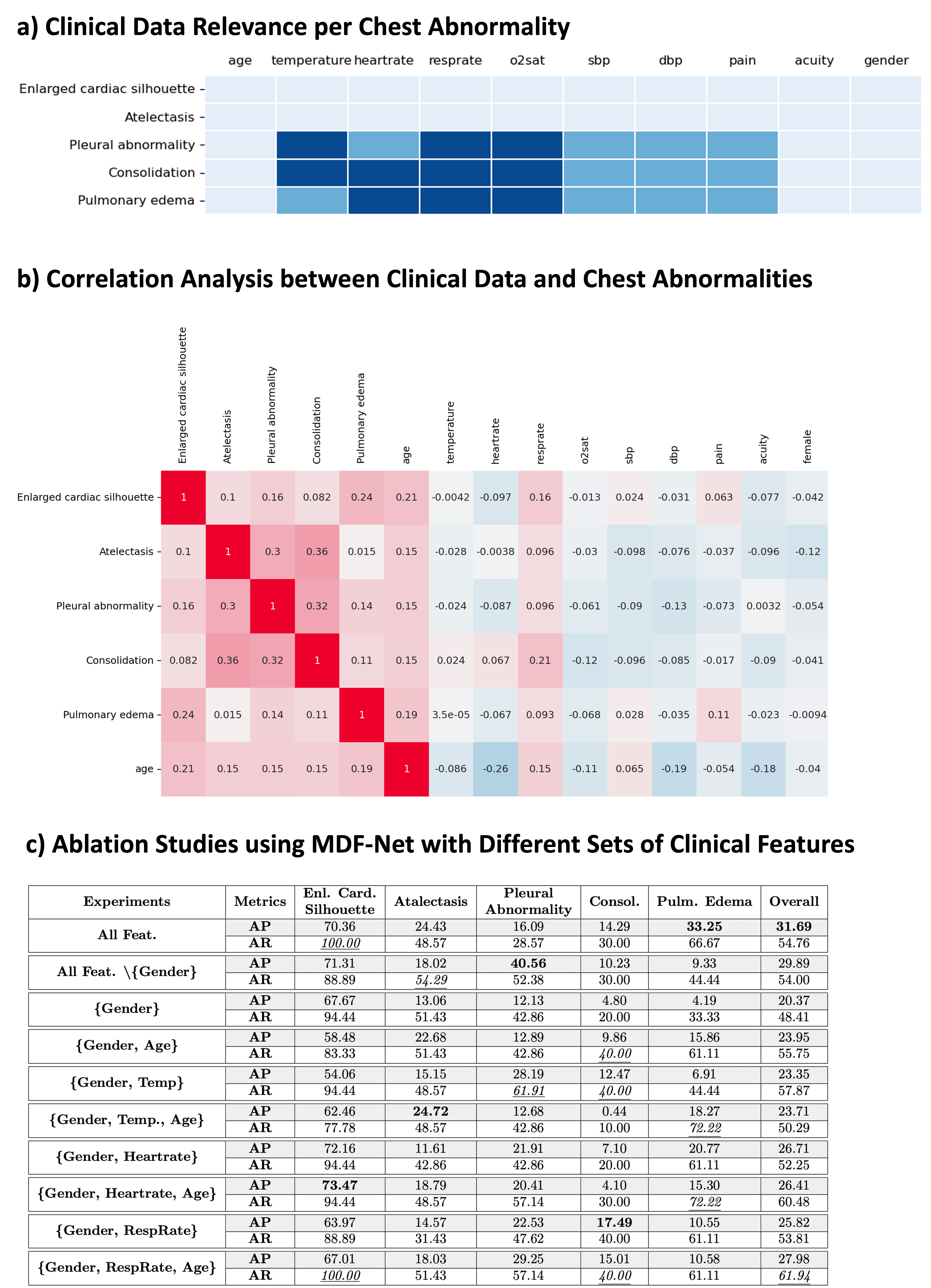}
    \caption{Panel \textbf{a)} The importance of clinical features for detecting each abnormality as reported by our radiology partners from Lus\'{i}adas Knowledge Center. Seven clinical features are highlighted for detecting pleural abnormality, consolidation, and pulmonary edema. Four out of seven highlighted features are considered the most significant. Panel \textbf{b)} Correlation matrix between clinical features and abnormalities. Panel \textbf{c)} Results of several ablation studies with different sets of features. Bold values indicate the best performing model in terms of AP, and the underlined values indicate the best performing model in terms of AR.}
    \label{fig:ClinicalDataAblationStudies}
\end{figure}

We also investigated the impact of different sets of clinical features in the proposed MDF-Net architecture. We first used radiologists' expertise to understand how the different features affect each chest abnormality (Figure \ref{fig:ClinicalDataAblationStudies} panel \textbf{a)}). The radiologists agreed that features such as body temperature are highly relevant for the indication of certain diseases that provoke infections, such as consolidation or pleural abnormality. We also investigated if this domain knowledge is present in the dataset by making a correlation analysis between the abnormalities present in each patient and the corresponding clinical data (Figure \ref{fig:ClinicalDataAblationStudies} panel \textbf{b)}). Finally, we applied the proposed MDF-Net and tested it on different sets of clinical features to investigate their impact on the learning model (Figure \ref{fig:ClinicalDataAblationStudies} panel \textbf{c)}).\\

\section*{Discussion}
\begin{figure}[!h]
    \centering
    \includegraphics[scale=0.5]{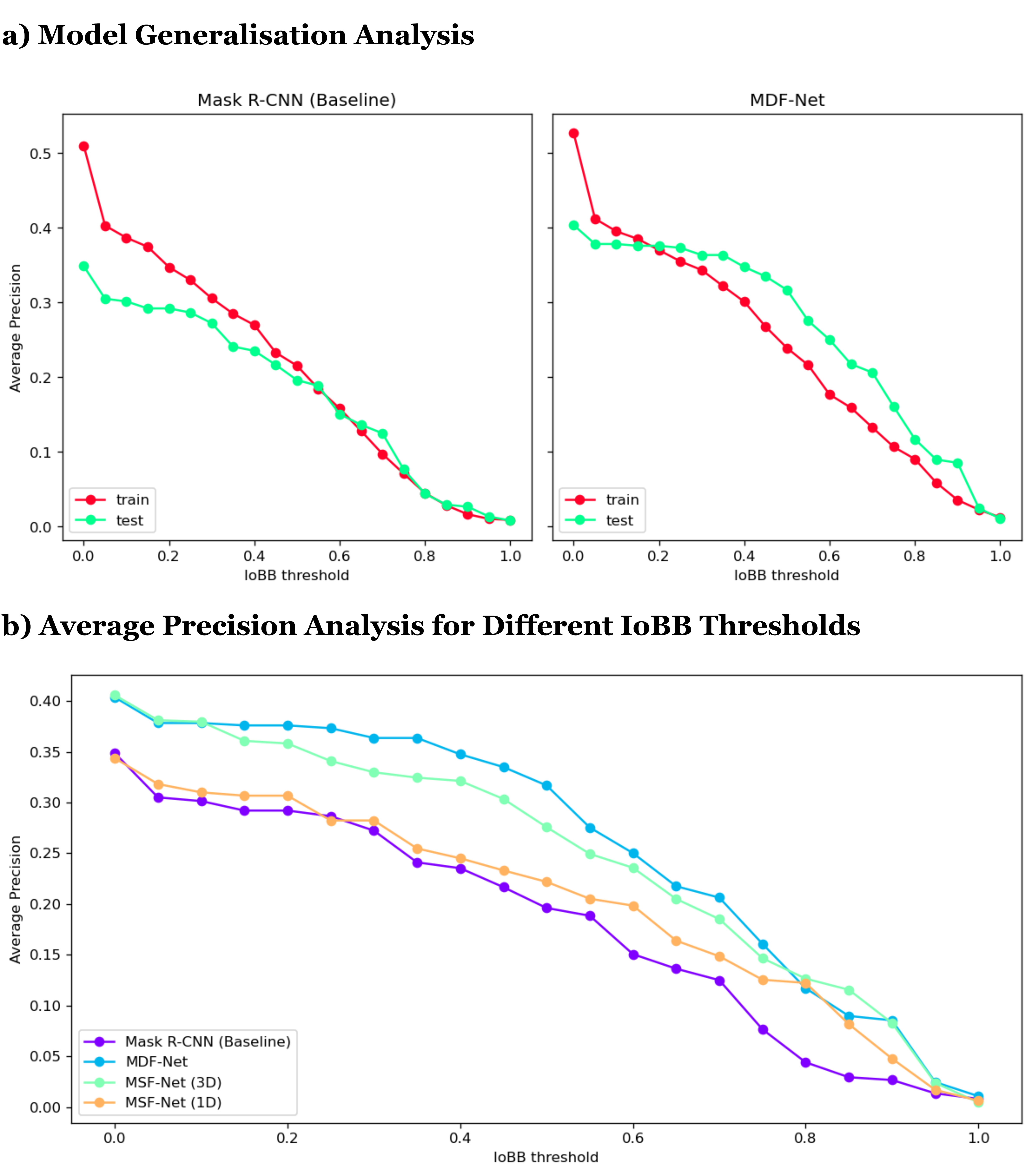}
    \caption{Panel \textbf{a)} Model generalisation analysis. From this graph, we show that clinical data and MDF-Net improve the generalization ability on the test set. Panel \textbf{b)} Average precision analysis for different IoBB thresholds: This chart shows MDF-Net reached the best performance when using both fusion methods. Only using 1-D or 3-D fusion methods alone can improve the model among all IoBB thresholds.
    }
    \label{fig:EvaluationIoBBs}
\end{figure}

The empirical evidence presented in the preceding section allows us to formulate the subsequent significant findings related to the proposed MDF-Net:

\begin{enumerate}
    \item MDF-Net outperforms the Mask R-CNN (Baseline) on overall AP and AR. MDF-Net has better AP on 4 out of 5 lesions and only has a small difference (-0.17\%AP) from the baseline model in detecting \textit{Pleural Abnormality}. In terms of overall performance, MDF-Net yields an improvement of +12\% AP and +0.85\% AR. 
    
    \item Figure \ref{fig:EvaluationIoBBs} panel \textbf{a)} shows the model performance on the training set (red) and test set (green). Although both models reached similar performance on the training set, MDF-Net generalised better to the test set compared to Mask R-CNN (Baseline). 
    
    \item In Figure \ref{fig:EvaluationIoBBs} panel \textbf{b)}, MDF-Net (blue) and the other two MSF-Nets outperform Mask R-CNN (purple) on almost all IoBB thresholds. While the Mask R-CNN (Baseline) suffers at a higher IoBB threshold standard, MDF-Net still maintains a reasonable performance, which indicates that MDF-Net can better locate lesions regardless of the IoBB threshold.
\end{enumerate} 

From the three findings above, we can experimentally prove that clinical data is of crucial importance and, thus, move away from the idea that using images alone can indicate reliable medical diagnosis.

\noindent\textbf{Ablation Study for fusion methods.} The Average Precision (AP) and Average Recall (AR) of each model are shown in Figure~\ref{fig:EvaluationFigure} panel~\textbf{c)}. 
In the previous subsection, we demonstrated the effectiveness of MDF-Net, and in this section, we use MSF-Net (1D) and MST-Net (3D) to conduct an ablation study. This experiment allowed us to test the effectiveness of each fusion method. The results are as follows:

\begin{enumerate}
    \item Figure~\ref{fig:EvaluationFigure} panel~\textbf{c)} shows MSF-Net (3D) has a larger improvement in performance (+7.96\%AP, +6.47\%AR) than MSF-Net (1D), which indicates the 3-D fusion is more effective than 1-D fusion. Compared to Mask R-CNN (Baseline), MSF-Net (3D) has better performance on 4 out of 5 lesions, which is the same as MDF-Net. And, We also noticed this model has the highest AR compared to all other models.
    \item The MSF-Net (1D) has a slight improvement on AP by +2.58\% but loss -4.508\% on AR, as shown in Figure \ref{fig:EvaluationFigure} panel \textbf{c)}. In MDF-Net (1D), the clinical data are only used to be concatenated with flattened RoIs, which means the clinical data are not involved in deciding where the Regions of Interest are (RPN output). In other words, if we split Mask-RCNN into two stages, the first stage determines the regions of interest (RoIs); and the second stage identifies the lesions inside those RoIs. In 1-D fusion, the clinical data is only perceived by the second stage, so the clinical data is not used for identifying RoIs. Consequently, this 1-D fusion only helps the final classifier to filter out regions misclassified by RPN; hence, AP increased while AR decreased.

    \item When using both 1-D and 3-D fusions together, the 3-D fusion can help RPN to pick up abnormal regions better while 1-D fusion can help the final classifier to determine whether a lesion exists in the given region.

\end{enumerate}

\noindent\textbf{Ablation Study for Clinical Features}. In order to understand the contribution and significance of clinical features, we also conducted an ablation study by giving a different combination of clinical features to MDF-Net. The performance of different combinations is shown in Figure \ref{fig:ClinicalDataAblationStudies} panel \textbf{c)}. When comparing the ablation result with the necessity table (Figures \ref{fig:ClinicalDataAblationStudies} panel \textbf{a)} and \ref{fig:ClinicalDataAblationStudies} panel \textbf{c)} and correlation matrix (Figure \ref{fig:ClinicalDataAblationStudies} panel \textbf{b)}, we found:

\begin{itemize}
    \item In Figure \ref{fig:ClinicalDataAblationStudies} panel \textbf{a)} radiologists stated that \textit{heartrate} is less important than \textit{temperature} and \textit{resprate} in diagnosing a pleural abnormality. In Figure  \ref{fig:ClinicalDataAblationStudies} \textbf{c)}, we found 
        \[
            AP_{\text{gender,heartrate}} < AP_{\text{gender,resprate}} < AP_{\text{gender, temperature}}\text{,}
        \] 
        which follows the importance shown in the necessity table (Figures \ref{fig:ClinicalDataAblationStudies} panel \textbf{a)}). However, when we introduce the \textit{age} feature, we obtain the following inequalities,
        \[
        AP_{\text{gender,age,temp}} < AP_{\text{gender,age,heartrate}} < AP_{\text{gender,age,resprate}}\text{,}
        \]
        which indicates that \textit{temperature} and \textit{resprate} did not bring more improvement to the model compared to \text{heartrate} when age is also used.\\

    \item In terms of diagnosing pulmonary edema, radiologists consider \textit{temperature} less important than \textit{heartrate} and \textit{resprate}. The same pattern as diagnosing pleural abnormality is shown.  \textit{Heartrate} and \textit{resprate} have higher AP than \textit{temperature} when \textit{age} is not used.
    
    \item Considering the effect of the feature \text{age}, $(\text{gender},\text{heartrate})$ is the only combination that has a slight performance drop (-0.298\% AP) when \textit{age} is introduced. The \textit{age} improves the ability to detect Atelectasis but damages the performance of diagnosing Consolidation. In terms of overall performance, \textit{age} gains improvement in most of the combinations, which is the same pattern shown in Figure \ref{fig:ClinicalDataAblationStudies} panel \textbf{b)} that \textit{age} has a higher correlation to most of the abnormalities.
    
    \item Moreover, in the same correlation matrix, we noticed the correlation between Consolidation and \textit{resprate} is higher than \textit{heartrate} and \text{o2sat}, and the ablation results also show \textit{resprate} improves models the most. Lastly, although \textit{resprate} has a high correlation with enlarged cardiac silhouette, the feature \textit{heartrate} seems more important in determining this abnormality.

\end{itemize}

\section*{Conclusion}
In this paper, we proposed a novel multimodal deep learning architecture, MDF-Net, and two fusion methods for multimodal abnormality detection, which can perceive clinical data and CXR images simultaneously. In MDF-Net, a spatialisation module is introduced to transform 1-D clinical data to 3-D space, which allows us to predict proposals with multimodal data. And, 1-D fusion is also used to provide clinical information to the final classifier in a residual manner. To test the performance of MDF-NET, we also propose a joining strategy to construct a multimodal dataset for MIMIC-IV. The experiments show that MDF-Net consistently and considerably outperforms the Mask R-CNN (Baseline) mode. And, both fusion methods show significant improvements in Average Precision (AP) while applying them together can achieve the best performance.

Overall, our MDF-Net improves upon the baseline Mask R-CNN by not only enhancing its ability to localize diseases in chest X-ray images but also extending its capabilities to incorporate vital clinical context, which is at the moment an important missing ingredient in the Deep Learning literature. This results in a more comprehensive and accurate diagnostic tool that can better support healthcare professionals and provide a better rationale for subsequent interpretations of the models.

In the future, we will explore the following two main directions:
    In this work, we can only retrieve 670 instances for our multimodal dataset, which is considered small compared to other popular datasets used for X-ray diagnosis. If there are larger-scaled datasets with clinical data available in the future, our work should also be tested on them to have a more objective evaluation.
    Our fusion methods can also be applied to other models, such as YOLO\cite{Redmon2015YOLO}, SSD\cite{Liu2016SSD} and DETR\cite{Carion2020DETR}. We will incorporate other architectures to evaluate the effectiveness of our fusion methods.

\section*{Methods}
The methods employed in this research comprise a combination of image processing and machine learning techniques to achieve effective disease detection in chest X-ray images. We have chosen a robust and well-tested algorithm as our foundation and modified it to better suit our specific objectives. This has led to the development of an innovative model that takes advantage of the synergy between traditional image data and structured clinical data. This combined use of data sources significantly enhances the model's diagnostic performance by providing more contextual information for accurate disease localization. The following subsection provides a detailed explanation of the model architecture used in this study.

\subsection*{Model Architecture}
 In this paper, we propose an extension of Mask R-CNN, MDF-Net. We chose Mask R-CNN as our baseline model for the following two reasons. First,  Mask R-CNN is a simple state-of-the-art model for object detection and instance segmentation, with proven and established success in localizing and classifying objects within images in a variety of contexts (see for instance \cite{Schweitzer18MaskRCNN,Liu19MaskRCNN,Conrady22MaskRCNN} that testifies its success with a wide range of applications). Given our task of disease localization in chest X-ray images, Mask R-CNN's ability to provide both the class and location of disease indications made it a suitable choice. Furthermore, Mask R-CNN's flexible and modular structure  (that is, easy to train) allowed us to extend and modify it to better suit our specific task and incorporate our novel elements.

The main innovation in our MDF-Net is the dual-fusion architecture that allows for the integration of both image and structured clinical data (i.e., tabular data). This is a significant departure from traditional Mask R-CNN models, which primarily work with image data alone or combine image and text data. By integrating clinical data, our model is able to consider the additional context that is crucial for accurate disease localization, thus making it more aligned with the actual diagnostic process of radiologists. Our model also includes a novel spatialization strategy, which converts clinical data into a 'pseudo-image' format that can be processed by the same convolutional layers as the image data. This is a significant innovation that allows for a more seamless and effective integration of the two data types.

\begin{figure}[!h]
    \resizebox{!}{\columnwidth}{
    \includegraphics{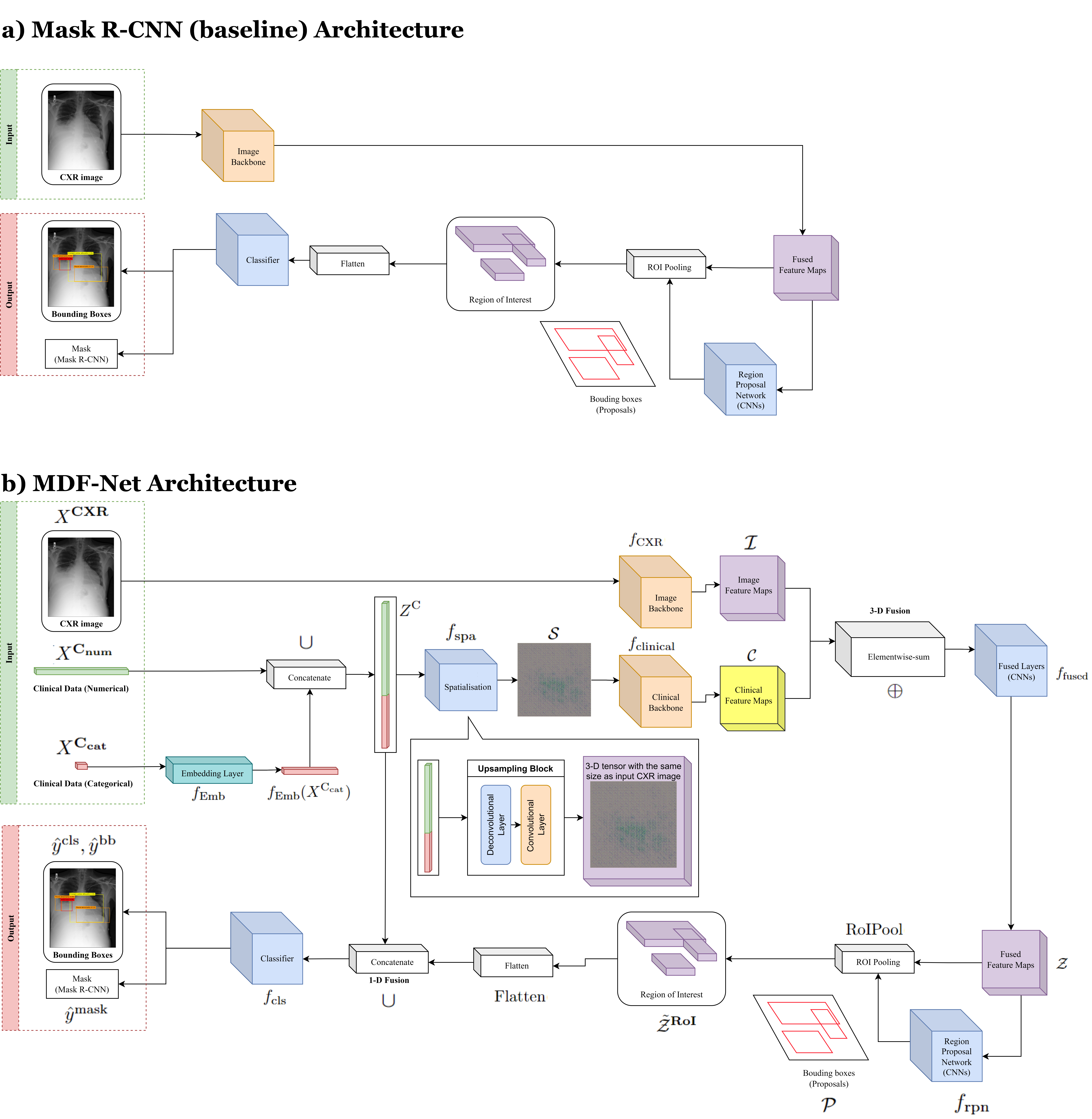}
    }
    \caption{Panel \textbf{a)} Mask R-CNN architecture. Panel \textbf{b)} Our proposed architecture and fusion methods.}
    \label{fig: ModelArchitectures}
\end{figure}

Figure~\ref{fig: ModelArchitectures} panel~\textbf{a)} shows the architecture of the original Mask R-CNN, which is the baseline model used in this work. The backbone can be any neural network that can extract feature maps from an image. The architecture of MDF-Net is shown in Figure \ref{fig: ModelArchitectures} panel \textbf{b)}. Considering the size of dataset, a small pre-trained backbone model, MobileNetv3 \cite{Howard2019MobileNetV3}, is used in both Mask R-CNN (baseline) and MDF-Net to prevent overfitting.

\textbf{Input Layer.} The proposed network receives as input two different modalities: a front (AP or anterior-posterior view) view of CXR images and the respective clinical data. They are defined as:

\begin{itemize}
    \item a set of CXR images: $X^{\text{CXR}} \in \mathbb{R}^{W \times H \times C}$;
    \item a set of clinical data (numerical features): $X^{\text{C}_{\text{num}}} \in \mathbb{R}^{n_1 \times 1}$;
    \item a set of clinical data (categorical features): $X^{\text{C}_{\text{cat}}} \in \{0,1\}^{n_2 \times 1}$.
\end{itemize}
In the proposed architecture, we set the dimensions of our image input space to $W=H=512, C=1$ (since the CXR image is gray-scale). In terms of the clinical data, our input space corresponds to the number of features: $n_1 = 9$ (continuous variables) and $n_2=1$ (categorical variable: gender). 

Since the input contains modalities with different dimensions, in order to perform fusion, we need to first implement feature engineering in this data to have the same dimensions before attempting any localized object detection learning.

\textbf{Feature Engineering:} $\{ X^{\text{C}_{\text{num}}}, X^{\text{C}_{\text{cat}}} \} \rightarrow \mathcal{C} \in \mathbb{R}^{W' \times H' \times D'}$

\text{~~~~~~~~~~~~~~~~~~~~~~~~~~~~~}$X^{\text{CXR}} \rightarrow \mathcal{I} \in \mathbb{R}^{W' \times H' \times D'}$

The goal of this step is to transform both the image input data and the clinical data to the same shapes. To achieve this, we do the following:

\begin{itemize}
    \item[]  \textbf{1. Image transform:} we extracted feature maps from $\mathcal{I}$ using a CNN backbone, $f_{\text{CXR}}$, which in this case corresponds to MobileNetv3 \cite{Howard2019MobileNetV3}:
    \[\mathcal{I} = f_{\text{CXR}}(X^{\text{CXR}}), \]
    where the resulting dimensional space is $\mathcal{I} \in \mathbb{R}^{W' \times H' \times D'}$. In our implementation, MobileNetv3 produced the features maps  $W'=H'=16$, $D' = 64$. We recognise that MobileNetv3 makes a significant reduction of our feature space, however, when we tried other CNN backbones such as ResNet, we obtained worse results and overfitting. MobileNetv3 provided the best results in our preliminary tests, hence our reason for choosing this backbone.
    
    \item[] \textbf{2. Clinical data encoding:} In terms of the feature maps for clinical data, the goal is to concatenate both numerical and categorical feature representations. To do so, we applied an embedding layer to the categorical features, $f_\text{Emb}$, so as to obtain a latent vector with the same dimensions of the numerical data. Next, we concatenate the resulting latent vector $f_\text{Emb}(X^{C_{cat}})$ with $X^{C_{num}}$ as follows:
    \[ Z^{\text{C}} = f_{\text{Emb}}(X^{\text{C}_{\text{cat}}}) \cup X^{\text{C}_{\text{num}}}\text{, where } Z^{\text{C}} \in \mathbb{R}^{n \times 1}, \text{~~~with } n = n_1 + n_2, \]
    where in our implementation $n=64$, $n1$ corresponds to the number of continuous features ($n_1=9$) and $n_2$ to the number of categorical features after being processed by an embedding layer ($n_2=55$); $\cup$ is the vector concatenation operation. However, the dimensionality $\mathcal{C}$ is different from the generated image feature maps, $\mathcal{I}$. This required an operation of transforming the dimensions of the clinical data from $n \ times $ to $W' \times H' \times D'$. To achieve this, we propose a method of spacialisation of the clinical data.
    
    \item[] 3. \textbf{Spatialisation}: We define a spatialisation layer, $f_{\text{spa}}$, as a deconvolutional layer \cite{Zeiler2010Deconv} followed by a convolution operation. The deconvolution takes as input the $n\times1$ dimensional clinical data vector and learns an upscaled representation using a sparse encoded convolution kernel \cite{Dumoulin2016CNNArithmetic, Albawi2017UnderstandingCNN}. This is given by 
    
    \begin{equation}    
         \mathcal{S} = f_{\text{spa}}(Z^C) = f_{L_\text{spa}}^{e}(Z^C), 
    \end{equation}
    \noindent where $f_{L_{\text{spa}}}(Z^C) = f_{\text{conv}}({f_{\text{deconv}}(Z^C)})$, and $e$ is the number of spatialised layers (in this work, we set $e=9$). After applying spatialisation, the size of $\mathcal{S}$ will become $W \times H \times C$, which is the size of input image $X^{\text{CXR}}$. 

    The primary purpose of using deconvolution, also known as transposed convolution, in this context, is to facilitate the upsampling of clinical data to a dimensionality that aligns with the input chest X-ray images. Achieving parity of dimensions between these two data types is crucial as it enables us to acquire feature maps of the same size from both datasets, thereby paving the way for a more effective fusion operation in the subsequent stages of our model.
    
    One of the significant benefits of deconvolution layers within this process is that they render the upsampling procedure trainable. In effect, this means that our model can learn the most efficient transformation strategy for converting the clinical data into a spatial 'pseudo-image', thereby enhancing its ability to integrate with the chest X-ray images and learn from the data concurrently.

    \item[] 4. \textbf{Clinical data transform}: 
    The last step for clinical data is to extract the feature maps from spatialised clinical data $\mathcal{S}$. By applying a CNN $f_{\text{clinical}}$, which uses the same architecture as $f_{\text{CXR}}$, we can obtain the clinical feature maps by: 
    \[
        \mathcal{C} = f_{\text{clinical}}(\mathcal{S}).
    \]
    
    In the end, with the proposed spatialisation operation $f_{\text{spa}}$ and the following CNN $f_{\text{clinical}}$, the resulting $\mathcal{C} \in \mathbb{R}^{W' \times H' \times D'}$, which matches the dimensions of $\mathcal{I}$. From this step, one can proceed to the fusion of both modalities. 
\end{itemize}

\noindent
\textbf{Input 3D Fusion: $\{ \mathcal{I}, \mathcal{C} \} \rightarrow \mathcal{Z} \in \mathbb{R}^{W' \times H' \times D'}$ }

The final feature map $\mathcal{Z}$ representing the element-wise sum fusion of both modalities is obtained by 
\begin{equation}
    \mathcal{Z} = f_{\text{fused}}(\mathcal{C} \oplus \mathcal{I}) \text{, }z \in \mathbb{R}^{H' \times W' \times D'},
\end{equation}

\noindent where $f_{\text{fused}}$ is another CNN module used for obtaining the features maps from the fused modalities, and $\oplus$ corresponds to the element-wise sum operation that is used for fusion. The final $\mathcal{Z}$ corresponds to the 3D feature map representation of the combined patient information. Next, we use this data representation as input to a Mask-RCNN architecture to perform abnormality detection.

\noindent
\textbf{Region Proposal Network: $\{ \mathcal{Z} \} \rightarrow \tilde{\mathcal{Z}}^{\text{RoI}} \in \mathbb{R}^{W_r \times H_r}$ }

To perform localized abnormality detection, we use the Region Proposed Network, $f_{\text{rpn}}$, of Mask-RCNN architecture to generate candidate object bounding boxes also known as proposals $\mathcal{P}$ given by
\begin{equation}
    \mathcal{P} = f_{\text{rpn}}(\mathcal{Z}) \text{, } \forall p_i \in \mathcal{P}: p_i = (x_i, y_i, w_i, h_i, c^{\text{obj}}_i).
\end{equation}
\noindent
RPN learns the coordinates of the generated bounding boxes $(x_i, y_i, w_i, h_i)$, and the corresponding confidence score, $c_{\text{obj}}$, of having an abnormality (object) in the localization of the bounding boxes. This confidence score is used to sort the generated proposals by their predictive relevance.

Using the coordinates of the computed bounding boxes, a RoIPool operation is performed to extract the corresponding Regions of Interest (RoIs), $\tilde{\mathcal{Z}}^{\text{RoI}} = \text{RoIPool}(\mathcal{P}, \mathcal{Z})$.
The RoIs result in a data structure with dimensions $\tilde{\mathcal{Z}}^{\text{RoI}} \in \mathbb{R}^{W_r \times H_r}$, where $W_r$ and $H_r$ are hyper-parameters. In our experiments, we set $W_r$ and $H_r$ to $7$.

\noindent
\textbf{Output: $\{Z^C, \tilde{\mathcal{Z}}^{\text{RoI}}\}  \rightarrow \hat{y}$}: 

After learning the candidate RoIs, we flatten this data to serve as input to a normal dense neural network, which will perform the final classification. In order to emphasize the role of the clinical data in this classification process, we concatenate the clinical data representation, $Z^C$, with the flattened candidate RoIs, $\tilde{\mathcal{Z}}^{\text{RoI}}$, before classification takes place. 
The role of the 1-Diffusion in the MDF-Net is to provide residual information to further pass the clinical data to deeper layers in our architecture. The final prediction $\hat{y}$ is then obtained by: 
\begin{equation}
    \hat{y} = f_{\text{cls}}(\text{Flatten}(\tilde{\mathcal{Z}}^{\text{RoI}}) \cup Z^{\text{C}}),
\end{equation}
where $\cup$ represents the vector concatenation operation, $\tilde{\mathcal{Z}}_{\text{RoI}} = \text{RoIPool}(\mathcal{P}, \mathcal{Z})$, and $\hat{y}$ contains predicted classes $\hat{y}^{\text{cls}}$, bounding boxes $\hat{y}^{\text{bb}}$, binary masks $\hat{y}^{\text{mask}}$, and $f_{\text{cls}}$ is the final classification layer. 

\noindent
\textbf{Number of Classes.} In this study we make object detection over five different classes: Enlarged Cardiac Silhouette, Atelectasis, Consolidation, Pleural Abnormality, and Pleural Edema. We chose to focus on five classes because they were the most representative in our dataset. We also took into account the constraints imposed by the dataset's class imbalance. The classes we have not included had insufficient examples, which would have posed significant challenges in training and evaluating our deep learning model.
Training a deep learning model with insufficient data for some classes could lead to overfitting and poor generalization performance for those classes. Moreover, it could bias the model towards the classes with more data. Therefore, to ensure a reliable and robust model, we chose to focus on the five most representative classes.

\subsection*{Model Complexity Analysis}

The overall computational complexity of the proposed architecture approximates the original Mask R-CNN model, which corresponds to the sum of the complexities of its components:
\[  O(NHW) + O(NCHW) + O(NCHW) \approx O(NCHW), \]
where, $N$ is the number of region proposals, $C$ is the number of classes (five classes, in our case), $H$ is the feature map height, and $W$ is the feature map width

The Faster-R CNN is composed of 5 main parts as follows: (1) a deep fully convolutional network, (2) region proposal network, (3) ROI pooling and fully connected networks, (4) bounding box regressor, and (5) classifier. 

The deep fully convolution network consists of five convolution layers, that is based  on Zeiler and Fergus’s \textit{fast} (smaller) model~\cite{Zeiler14CNNs}. Having an image $I$, this step extracts $256 \times N \times N$ feature maps  (given the 5 convolution layers~\cite{Zeiler14CNNs}). This is the input of the RPN network and ROI pooling layer.  In the RPN network, for each point of the feature map, there are, say K, anchors (or candidate window, typically 2000) with different scales and rations. Thus,  there will be a total of  N x N x K candidate widows. Non-maximum suppression allows to obtain typically $2000$ \cite{Ren15FastRCNN} candidate windows. This yields a complexity of  $O(N^2)$.

Using the candidate windows and the feature map above, RoI pooling layer divides the varied size candidate windows into an $H \times W $ grid of sub-windows then max-pooling the values in each sub-window into the corresponding output grid cell. The complexity of this process is $O(1)$.

\subsection*{Training}

Once the model architecture has been established, the next crucial step involves training the model using a carefully designed loss function to achieve optimal performance. In the context of Mask R-CNN, the loss function plays a pivotal role in learning the optimal parameters of the model. A well-constructed loss function balances multiple objectives, including accurate classification of abnormalities, precise bounding box regression, and efficient object proposal.

In our training process, we incorporate five loss terms, each addressing a specific aspect of the model's learning objectives. These loss terms aim to guide the model toward achieving high precision in identifying and localizing diseases in chest X-ray images. In the following, we provide a detailed explanation of each of these loss terms.

\begin{itemize}
    \item $L_{\text{cls}}$: Cross-entropy between groundtruth abnormality $y^{\text{cls}}$ and predicted abnormalities $\hat{y}^{\text{cls}}$. This loss term requires the model to predict the class of abnormalities correctly in the output layer.
    \item $L_{\text{bb}}$: Bounding box regression loss between ground-truth bounding boxes $y^{\text{bb}}$ and predicted bounding boxes $\hat{y}^{\text{cls}}$ calculated using smooth-L$_1$ norm:
        \begin{equation}
            L_{bb} =  \sum_{i=1}^{n} l_i\text{, where } l_i = \begin{cases}
            0.5(\hat{y}_{i}^{\text{bb}} - y_{i}^{\text{bb}})^2/\beta &, \text{if $\hat{y}_{i}^{\text{bb}} - y_{i}^{\text{bb}} < \beta$ }\\
            \lvert \hat{y}_{i}^{\text{bb}} - y_{i}^{\text{bb}} \rvert  - 0.5*\beta &, \text{otherwise}
            \end{cases},
        \end{equation}
        \noindent $\beta$ is a hyperparameter. In our implementation, $\beta = \frac{1}{9}$. To minimise this loss, the model has to locate abnormalities in the correct areas in the output layer.
        
    \item $L_{\text{mask}}$: Binary cross-entropy loss between ground-truth segmentation $y^{\text{mask}}$ and predicted masks $\hat{y}^{\text{mask}}$, which requires the model to locate abnormalities at the pixel level.
    \item $L_{\text{obj}_{\text{rpn}}}$: Binary cross-entropy loss between ground-truth objectness $y^{\text{obj}}$ and predicted objectness $c^{\text{obj}}$ (confidence score), which requires RPN to correctly classify whether the proposals (candidate bounding boxes) contain any abnormality.  
    \item $L_{\text{bb}_{\text{rpn}}}$: Proposal regression loss between proposals (candidate bounding boxes) $p: p \in \mathcal{P}$ and ground-truth bounding boxes $y^{\text{bb}}$, which is also calculated using the same smooth-L$_1$ norm function for $L_{\text{bb}}$. This loss term aims to improve RPN on localising abnormalities.   
\end{itemize}

We used homoscedastic (task) uncertainty \cite{Kendall2017TaskUncertaintyLossWeighting} to train the proposed model using these five loss terms by dynamically weighting them for better convergence. Let $\mathcal{L} = \{L_{\text{cls}}, L_{\text{bb}}, L_{\text{mask}}, L_{\text{obj}_{\text{rpn}}}, L_{\text{bb}_{\text{rpn}}} \}$, we used SGD (stochastic gradient descent) to optimise the overall loss function

\[
    \arg\min_{\theta, \alpha_l} \sum_{l \in \mathcal{L}}\frac{1}{2\alpha_{l}^{2}}l(\theta)+\log{\alpha_{l}^{2}},
\]

\noindent where $\theta$ is the wights of MDF-Net, and $\alpha_{l}$ is a trainable parameter to weigh each task/loss. 

The validation of this architecture required a multimodal dataset. In this study, we used medical data, more specifically chest X-ray images from MIMIC-CXR \cite{Johnson2019MIMIC_CXR} and patient's clinical data from MIMIC-IV-ED \cite{Johnson2021MIMIC_IV_ED}. However, these datasets are offered separately in the literature, and a thorough data integration had to be conducted before evaluating our architecture.

\subsection*{Dataset}

Modern medical datasets integrate both imaging and also tabular data. The latter refers to  medical history and lifestyle questionnaires, where clinicians have the responsibility to combine the above two sources of information. Note also that beyond diagnosis, multimodal data  (i.e., comprising tabular and image data) is crucial to the advance and understanding of diseases motivating the creation of the so-called biobanks. There are several examples of biobanks, for instance, German National Cohort~\cite{Germandata14} or the UK Biobank~\cite{Sudlow15UK} that includes thousands of data fields from patient questionnaires including data from questionnaires, physical measures, sample essays, accelerometry, multimodal imaging, genome-wide genotyping. However, these datasets do not contain local image annotations of lesions. Therefore, we propose a strategy to combine MIMIC-IV~\cite{Johnson2021MIMIC_IV} and REFLACX~\cite{Lanfredi2021REFLACX} to create our dataset, MIMIC-Eye, from scratch that meets the requirement of this work which can be accessed in physionet \cite{MIMIC-EYE23}. The Medical Information Mart for Intensive Care (MIMIC) IV dataset is from two in-hospital database systems, a custom hospital-wide EHR and an ICU-specific clinical information system, in Beth Israel Deaconess Medical Center (BIDMC) between 2011 and 2019. The MIMIC-IV database is grouped into three modules, including \textit{core}, \textit{hosp}, and \textit{icu}. In this work, only the patients' data in the \textit{core} module is used. 

As well as the MIMIC-IV dataset, two other MIMIC-IV subsets, MIMIC-IV ED (Emergency Department) \cite{Johnson2021MIMIC_IV_ED}, and MIMIC-IV CXR (Chest X-ray) \cite{Johnson2019MIMIC_CXR}, are  used to create the multimodal dataset. These two datasets can be linked to the MIMIC-IV dataset with \textit{subject\_id} and \textit{stay\_id}. The MIMIC-IV ED dataset was extracted from the emergency department at the Beth Israel Deaconess Medical Center. It contains data for emergency department patients collected while they are in the ED. The \textit{triage} data of the MIMIC-IV ED dataset is one source providing patients' health condition in this work, such as \textit{temperature}, \textit{heart rate}, \textit{resprate}, etc. MIMIC-CXR is another subset of MIMIC-IV consisting of 227,835 radiographic studies and 377,110 radiographs from BIDMC EHR between 2011 - 2016. In the original MIMIC-CXR dataset, the CXR images are provided in \textit{DICOM} format, which allows radiologists to adjust the exposure during reading. However, to train a machine learning model, the \textit{JPG} file is preferred. The author of MIMIC-IV CXR then presented MIMIC-CXR JPG dataset \cite{DJohnson2019MIMIC_CXR_JPG} to facilitate the training process. REFLACX dataset is another subset of MIMIC-IV ED, which provides extra data from different modalities, such as eye tracking data, bounding boxes, and time-stamped utterances. The bounding boxes in REFLACX are used as groundtruth in this work.

In total ten clinical features are used in this work. The MIMIC-IV Core \textit{patients} data includes only two clinical attributes, age and gender. And the other eight clinical features are extracted from the MIMI-IV ED \textit{triage} data. The explanations for these eight clinical features in the MIMIC-IV documentation are:

\begin{enumerate}
    \item \textbf{temperature}: The patient’s temperature in degrees Fahrenheit.
    \item \textbf{heartrate}: The patient’s heart rate in beats per minute.
    \item \textbf{resprate}: The patient’s respiratory rate in breaths per minute.
    \item \textbf{o2sat}: The patient’s peripheral oxygen saturation as a percentage.
    \item \textbf{sbp, dbp}: The patient’s systolic and diastolic blood pressure, respectively, measured in millimetres of mercury (mmHg).
    \item \textbf{pain}: The level of pain self-reported by the patient, on a scale of 0-10.
    \item \textbf{acuity}: An order of priority. Level 1 is the highest priority, while level 5 is the lowest priority.
\end{enumerate}

Before explaining the creation process, it is necessary to introduce some important IDs and data tables in the MIMIC-IV dataset. Four important IDs are used in MIMIC-IV to link the information across tables. They are:

\begin{itemize}
    \item \textbf{subject\_id (patient\_id)}: ID specifying an individual patient.
    \item \textbf{stay\_id}: ID specifying a single emergency department stay for a patient.
    \item \textbf{study\_id}: ID specifying a radiology report written for the given chest x-ray. It is rarely mentioned because we do not use the report as the groundtruth label in this paper.
    \item \textbf{dicom\_id}: ID specifying a chest x-ray image (radiograph).
\end{itemize}

And the following four tables in MIMIC-IV are used to create our multimodal abnormality detection dataset:

\begin{itemize}
    \item \textbf{MIMIC-IV Core patients}: Information that is consistent for the lifetime of a patient is stored in this table, including age and gender.
    \item \textbf{MIMIC-IV ED triage}: This table contains information about the patient when they were first triaged in the emergency department, including temperature, heart rate and more clinical data.
    \item \textbf{MIMIC-IV Core edstays}: Provides the time the patient entered the emergency department and the time they left the emergency department, which helps us to identify the \textit{stay\_id} for CXR images.
    \item \textbf{MIMIC-IV CXR metadata}: Contains the information about the CXR image (radiograph), including the time taken, height and width.
\end{itemize}

\section*{Limitations and Ethical Considerations}

Although our study presents promising results, several limitations should be acknowledged. As with all scientific research, our study contains inherent limitations, primarily revolving around data selection, fusion methods, and ethical considerations:

\begin{itemize}
    \item[] \textbf{Dataset Choice.} The effectiveness of our proposed model relies heavily on the quality and comprehensiveness of the input information, both images and clinical data. While using publicly available and well-established datasets such as MIMIC-CXR, MIMIC IV-ED, and REFLACX minimizes the risk of data quality issues, these datasets may only partially represent diverse global populations. The performance of our model could vary when applied to different demographic groups, and potential biases in the datasets could influence the results.

    \item[] \textbf{Dataset Size.} Although many public datasets contain both CXR images and manual lesion annotations, unfortunately, to the best of our knowledge, we are unaware of any dataset that also contains the patients' clinical data. Due to privacy concerns, most publicly available medical image datasets do not include this kind of clinical information. Patient clinical data are sensitive and protected by strict privacy regulations. As a result, researchers often face significant challenges in obtaining datasets that combine imaging data with relevant clinical information. This policy limits the effectiveness and ability of our MDF-Net to generalize.

    \item[] \textbf{Limitations in Fusion Methods.} For our study, we considered other fusion methods, such as the Laplacian pyramid and adaptive sparse representation \cite{Wang2020}, for the 3D fusion component in the proposed MDF-Net. However, these methods are not differentiable, which makes them incompatible with our end-to-end deep learning architecture. The backpropagation process used to train deep learning models requires the gradients (derivatives) of the loss concerning the model parameters. Non-differentiable operations disrupt this gradient flow, which could lead to suboptimal or untrainable models.
 
\end{itemize}

While the development and application of multimodal deep learning technologies have the potential to enhance disease diagnosis significantly, several ethical considerations must be addressed to ensure that such technologies are used responsibly and effectively.

\begin{itemize}
   \item[] \textbf{Impact on Healthcare Professionals.} While applying deep learning technologies may streamline diagnostic procedures and alleviate the workload of healthcare professionals, we must consider the potential impact on their roles and responsibilities. In this study, we align our work with the perspective that these technologies are tools that \textit{can support, not replace}, healthcare professionals. Our interviews with radiologists highlighted the importance of integrating clinical data in the image diagnosis process, emphasizing the continued need for expert knowledge and more human-centred deep learning architectures as proposed in this paper.

    \item[] \textbf{Impact on Trust.} The black-box nature of deep learning models raises concerns about transparency and trust in AI decisions. Our multimodal DL architecture seeks to improve the interpretability of predictions by using clinical information alongside image data, providing a context for the decisions made by the model. Further development of these technologies must emphasize explainability, so healthcare professionals and patients can understand and trust the diagnoses provided by these models. By providing lesion detection of the predicted lesions, we are already promoting one layer of interpretability. For future work, we are already developing methods to translate the symbolic representation of identified lesion bounding boxes into human-level explanations incorporating domain knowledge.

    \item[] \textbf{Biases.} Potential biases in the model's predictions, resulting from biased or unrepresentative training data, can lead to disparities in healthcare outcomes. Care must be taken to ensure that datasets used to train such models are representative of the diverse patient populations they will serve. In cases where data is imbalanced, techniques such as oversampling, undersampling, or synthetically augmenting the minority class should address this issue. However, this may reinforce potential selection biases in the dataset. For this reason, we restricted our study to the most frequent classes in the dataset (ending up in a much smaller dataset that impacted our model's performance, rather than introducing selection and sampling biases from data augmentations techniques.
\end{itemize}

\section*{Data Availability}
The code for this work is available on GitHub at \url{https://tinyurl.com/multimodal-abn-detection}.

\bibliography{sn-bibliography}

\section*{Acknowledgements}
This material is based upon work supported by the UNESCO Chair on AI\&XR; and the Portuguese \textit{Funda\c{c}\~{a}o para a Ci\^{e}ncia e a Tecnologia (FCT)} under grants no. 2022.09212.PTDC (XAVIER) and no. UIDB/50021/2020.

\section*{Author contributions statement}

\noindent
Chihcheng Hsieh: Conceptualisation, Methodology, Implementation, Mathematical Formulation, Evaluation, Writing 

\noindent
Isabel Nobre, Sandra Costa: Study idea and original insight, Clinical Knowledge, Review 

\noindent
Sandra Costa: Study idea and original insight, Clinical Knowledge, Review 

\noindent
Chun Ouyang: Supervision, Review, Writing 

\noindent
Margot Brereton: Supervision Review, Writing 

\noindent
Jacinto Nascimento: Conceptualization, Methodology, Writing, Review 

\noindent
Joaquim Jorge: Review, Writing 

\noindent
Catarina Moreira: Conceptualization, Methodology, Mathematical Formulation, Writing, Supervision, Review

\end{document}